\documentclass{article}

\usepackage[preprint]{neurips_2021}

\usepackage[utf8]{inputenc} 
\usepackage[T1]{fontenc}    
\usepackage{hyperref}       
\usepackage{url}            
\usepackage{booktabs}       
\usepackage{amsfonts}       
\usepackage{nicefrac}       
\usepackage{microtype}      
\usepackage{xcolor}         

\usepackage{amsmath}
\newtheorem{lemma}{Lemma}
\newtheorem{conj}{Conjecture}

\usepackage{graphicx}
\usepackage{subfig}

\usepackage{wrapfig}

\title{Neural Options Pricing}

\author{%
  Timothy DeLise\\
  Department of Mathematics and Statistics\\
  Université de Montréal\\
  \texttt{timothy.delise@umontreal.ca} \\
}

\begin{document}

\maketitle

\begin{abstract}
  This research investigates pricing financial options based on the traditional martingale theory of arbitrage pricing applied to neural SDEs. We treat neural SDEs as universal Itô process approximators. In this way we can lift all assumptions on the form of the underlying price process, and compute theoretical option prices numerically. We propose a variation of the SDE-GAN approach by implementing the Wasserstein distance metric as a loss function for training. Furthermore, it is conjectured that the error of the option price implied by the learnt model can be bounded by the very Wasserstein distance metric that was used to fit the empirical data.    
\end{abstract}

\section{Introduction}

Neural SDEs are an interesting new development at the crux of the fields of deep learning, stochastic processes, and differential equations \citep{tzen2019neural,tzen2019theoretical,peluchetti2020infinitely,pmlr-v108-li20i}. This article explores option pricing theory under neural SDEs. Indeed, financial applications are listed as a target application of neural SDEs in  \cite{pmlr-v108-li20i}. The notion that asset price processes follow SDEs (Itô processes) is one of the basic assumptions in modern mathematical finance theory \citep{steele2012stochastic, bjork}. For instance, the celebrated Black-Scholes formula assumes that asset prices follow geometric Brownian motion (2 degrees of freedom). Meanwhile, others have proposed that true market processes follow different, usually \textit{heavy-tailed}, distributions \citep{MANDELBROT20031, Rachev2005FatTailedAS, Mandelbrot1997}. However, these latter ideas often lack the mathematical simplicity, structure, and allure of the former, especially for popular applications, like options pricing. The advent of neural SDEs entices us to re-think traditional norms in the pursuit of parsimony and flexibility.

In this study we explore option pricing under the assumption that we can price processes are Itô processes that follow some SDE, and that we can approximated them to arbitrary accuracy with neural SDEs\citep{tzen2019theoretical}. This assumption is applied to the martingale theory of options pricing \citep{steele2012stochastic}. We then build a pipeline which computes the option price that is implied by the trained model. The benefit of this approach is that we have lifted any assumption on the specific form of the price process SDE.

Along the way we have implemented a variation of the SDE-GAN\citep{kidger2021neural} approach to train our neural SDEs to target data. The Wasserstein-1 distance \citep{villani2003topics} is implemented as a loss function to minimize the error between the target and model via gradient descent. This computation is outlined by \cite{ramdas2015wasserstein} for the 1-dimensional case, which currently limits our implementation to 1 dimension. However restrictive, we find to be quick, direct, and easily implemented using the \emph{torchsde} package \citep{pmlr-v108-li20i,kidger2021neural}. An upshot of this training method is a conjectured upper bound on the error of the options price implied by the trained neural SDE. We propose that the error in the computed options price is bounded by the loss (Wasserstein distance) attained by the neural SDE model during training.

Finally, we illustrate the complete pipeline for computing the price of a European call option in a toy example based on empirical samples drawn from geometric Brownian motion. We train a neural SDE, where both drift and diffusion functions are multi-layer perceptrons. We then compare our computed options price with the theoretical price which is given by the Black-Scholes formula\citep{bs}. At the same time we empirically investigate our conjectured bound. The experiment supports the claim that the options price is correctly computed and the error is bounded by the Wassterstein-1 loss value that was attained during training.

The following section gives some background in neural SDEs, the martingale theory of arbitrage pricing, and the Wasserstein distance metric. After that, \textit{neural options pricing} is presented, followed by the experiments. At the end there is a discussion of this technique, including its limitations, and directions for future work.

\section{Background Information}

\subsection{Neural SDEs}

Continuous-time models in deep learning have received an increasing amount of attention in recent years, highlighted by the success of neural ODEs \citep{chen2019neural}. A neural ODE can be described as a resnet \citep{he2015deep} with infinite depth. Since then, several researchers have worked on extending this idea to SDEs. \cite{tzen2019theoretical} address the idea of the continuum limit of deep generative models previously proposed by \cite{rezende2014stochastic}. Separate research by \cite{peluchetti2020infinitely} proposes imposing time on the layers of deep networks with diffusion, and draws a connection between stochastic processes and infinitely deep residual networks. In either case, the conclusion is that of a neural SDE: an SDE where the drift and diffusion functions are neural networks. In the present work, we will consider SDEs of this form, where $\mu_\theta$ and $\sigma_\phi$ are the neural network drift and diffusion processes parametrized by $\theta$ and $\phi$, which represent all the tune-able weights of each network. Equation \ref{eq:nsde} defines the stochastic process $S_t$ by way of a neural SDE. $B_t$ is a Brownian motion, and the entire process is adapted to the natural filtration generated by the random trajectories $B_t$. For more background in SDEs, see \cite{steele2012stochastic}.
\begin{equation} \label{eq:nsde}
    dS_t = \mu_\theta(t, \omega)dt + \sigma_\phi (t, \omega) dB_t
\end{equation}

It has been shown in \cite{tzen2019theoretical} that there exist neural SDEs that can approximate any target measure to arbitrary accuracy. Further studies have run with this idea and have described efficient methods for training such models. The authors of \cite{pmlr-v108-li20i} have developed an latent-SDE method for learning time-series using a latent neural SDE. Training methods are adapted from \cite{chen2019neural} which address the natural difficulties moving from ODEs to SDEs and scale efficiently in memory and time. More recently \cite{kidger2021neural} proposed training neural SDEs as generative models in a Wasserstein-GAN architecture \citep{arjovsky2017wasserstein}, which is an alternative to the latent-SDE mentioned above. In this framework the aim is to effectively minimize the Wasserstein distance between the target and generative model by way of a mini-max game involving a discriminator network. In this work we wish to assume that our price process is directly modeled by a neural SDE, so the latent-SDE approach is not appropriate. Instead we adapt a simple training scheme that is theoretically analogous to SDE-GAN\citep{kidger2021neural} in section \ref{sec:nop}.

\subsection{Martingale Theory of Arbitrage Pricing}
\label{sec_martingale}

This theory is a cornerstone of modern mathematical finance and many books have been written on the topic \citep{steele2012stochastic,bjork}. The core assumptions are that asset prices follow Itô processes, which are continuous processes over some probability space, adapted to the Brownian filtration and can be written as an SDE in \textit{standard form} like in equations \ref{eq:ito1} and \ref{eq:ito2} which are equivalent.
\begin{equation}\label{eq:ito1}
    S_T = S_t + \int_t^T \mu(s, \omega)ds + \int_t^T \sigma(s, \omega) dB_s
\end{equation}
\begin{equation}\label{eq:ito2}
    dS_t = \mu(t, \omega)dt + \sigma(t, \omega) dB_t
\end{equation}

Interested readers are invited to explore more about this basic assumption. The remaining core assumption of the theory is that arbitrage (risk-free profits) cannot exist in the market. The application here is to price \textit{derivatives} or \textit{contingent claims}, both of which simply refer to contracts that have an expiration date and are represented by simple math. Here we will only consider options contracts, and in particular,  European call options. These options are parametrized by a strike price $K$ and expiration date $T$ and operate on an underlying asset. The holder of the contract has the right to purchase the underlying asset at the strike price on the expiration date. 

Assuming our asset price follows the Itô process $S_t$ described in equation \ref{eq:ito1}, then the price of a European call option (let's call it $f$) at the expiration date $T$ is given by the following formula. 
\begin{equation}
    f(S_T) = \max(S_T - K, 0)
\end{equation}

The argument here is the key to the whole theory. If we are at time $T$, then the price of this contract must equal $f(S_T)$. If the price were different then there would be an arbitrage opportunity, where anyone could make an instantaneous risk-free profit.

This basic idea has far-reaching ramifications for the theory. Of course, no one is interested in buying option contracts at the time they expire, hence the point is to understand what the price should be at any time $t < T$. The upshot of the theory is that at any such time $t$, the price of any derivative contract (not just options) must be the expected value of the price of the contract at expiration over a \textit{risk-neutral equivalent martingale measure}. We will call the price of the European call option $V_t$, then
\begin{equation}
    V_t = \mathbb{E}_Q[f(S_T)]
\end{equation}

and $Q$ is this risk-neutral equivalent martingale measure over which we take the expectation. Here we are going extremely light on the math, since describing it all in full detail is far beyond the scope of this paper. The original, or \textit{natural}, probability measure of the process $S_t$ is refered to as $P$. The measure $Q$ is often understood by of the Radon-Nikodym derivative $M_t = \frac{dQ}{dP}$ which is defined for any standard process in the foundational work of \cite{girsanov}.
\begin{equation}
    M_t = \exp\left( \int_0^t \frac{\mu(s,\omega)}{\sigma(s,\omega)} dBs + \frac{1}{2} \int_0^t \left( \frac{\mu(s,\omega)}{\sigma(s,\omega)} \right)^2 ds \right) 
\end{equation}

Under Girsanov's theory, $B_t$ is no longer Brownian motion under $Q$. Instead $$\tilde{B}_t = B_t + \int_0^t \frac{\mu(s,\omega)}{\sigma(s,\omega)} ds$$ is Brownian motion under $Q$. Using this substitution we can then convert our price calculation back into the natural probability measure $P$, which gives a technique for computing the price numerically. For expectations under $P$ we drop the subscript.
\begin{align*}
    V_t &= \mathbb{E}_Q[f(S_T)]\\
    &= \mathbb{E}_Q \left[f \left( S_0 + \int_0^t \mu(s, \omega)ds + \int_0^t \sigma(s, \omega) dB_s \right) \right]\\
    &= \mathbb{E}_Q \left[f \left( S_0 + \int_0^t \sigma(s, \omega) \left( dB_s + \frac{\mu(s,\omega)}{\sigma(s,\omega)} ds \right) \right) \right]\\
    &= \mathbb{E} \left[f \left( S_t + \int_t^T \sigma(s, \omega) dB_s \right) \right]
\end{align*}

\subsection{Wasserstein Distance Metric}

The Wasserstein distance is a distance metric which measures the distance between probability measures over a metric space. It comes from optimal transport theory and is increasingly popular in statistics and machine learning \citep{villani2003topics,arjovsky2017wasserstein}. The Wasserstein distance is parametrized by an integer $p$ and the common definition is as follows. Let $M$ be a metric space for which every measure $M$ is a Radon measure and let $P_p(M)$ be the collection of probability measures on $M$ with $p^{\text{th}}$ moment, then the Wasserstein-$p$ distance between two measures $\mu$ and $\nu$ on $P_p(M)$ is defined as in equation \ref{eq:wass1}.
\begin{equation} \label{eq:wass1}
    W_p(\mu,\nu)=\left( \inf_{\gamma \in \Gamma(\mu,\nu)} \int_{M \times M} d(x,y)^p d\gamma(x,y) \right)^{1/p}
\end{equation}

Here, $\Gamma$ is the collection of all measures on $M \times M$ with marginals $\mu$ and $\nu$, also called couplings between $\mu$ and $\nu$. There are some interesting mathematics surrounding this equation. Firstly, in our experiment we are working with empirical distributions in $\mathbb{R}^d$ and $d=1$, then the Wassterstein distance can be directly computed by comparing the empirical cumulative distribution functions, as follows.
\begin{equation} \label{eq:wass2}
    W_p(\mu,\nu)=\left( \int_0^1 \bigg|F^{-1}(z) - G^{-1}(z) \bigg|^p dz \right)^{1/p}
\end{equation}

In particular, the optimal transport strategy $\inf_{\gamma \in \Gamma(\mu,\nu)}$ can be solved using linear programming techniques and yields a dual formulation of the problem. In this paper we will focus on the case when $p=1$, which is usually referred to as the \textit{earth-mover's distance}. In this case, the dual formulation yields the following expression.
\begin{equation} \label{eq:wass3}
    W_1(\mu,\nu)= \sup_{f \in \mathcal{F}} \left\{ \int f(x) d\mu(x) - \int f(x) d\nu(x) \right\}
\end{equation}

Where $\mathcal{F}$ is the family of all continuous functions with Lipschitz constant less than or equal to 1. If we are given a function $f \in \mathcal{F}$, then this previous equation is equivalent to writing the following.
\begin{equation} \label{eq:wass4}
    \mathbb{E}_{x \sim \mu}[f(x)] - \mathbb{E}_{x \sim \nu}[f(x)] \le W_1(\mu,\nu)
\end{equation}

\section{Neural Options Pricing}
\label{sec:nop}

The pieces are now in place to describe \textit{neural options pricing}, the main contribution of this article. The idea is straight-forward. We assume that we are given empirical trajectories (data) from a price process of an underlying asset. We call the target process $R_t$. We fit a neural SDE $S_t$ (equation (\ref{eq:nsde})) to the price process. The main assumption here is that $R_t$ follows a standard Itô process that we don't have access to and we are trying to approximate it with $S_t$. The mathematical gap we are trying to fill is to say, if we have approximated $R_t$ to a certain level of accuracy with $S_t$, then we have also approximated its options price to a similar level of accuracy. This is the main motivation behind using the Wasserstein-1 distance as a loss function of our training algorithm. Our option contract function $f$ belongs to the family $\mathcal{F}$ of equation \ref{eq:wass3} because it is a Lipschitz-1 function.

We consider the Wasserstein-1 distance between the empirical distributions of $R_t$ and $S_t$ in discrete time steps over many trajectories and use this as the loss for training via gradient descent methods. This collection of trajectories is considered a batch of training data. Using the technique of equation \ref{eq:wass2} as a loss function, we then use gradient descent optimization to update the parameters of the Neural SDE model. While our experiments show that this method produces good training performance, we also notice an added benefit. Since the contract function $f \in \mathcal{F}$, then we have the following inequality by definition in equation (\ref{eq:wass4}).
\begin{equation}\label{eq:nsde2}
 \bigg| \mathbb{E}[f(S_T)] - \mathbb{E}[f(R_T)] \bigg| \le W_1(S_T,R_T)
\end{equation}
In order to compute the options prices corresponding to $S_T$ and $R_T$, we need to apply the Girsanov change of measure to $S_T$ and $R_T$ and compute expectations. Let $W$ denote the risk-neutral martingale measure with respect to $S_T$ and $V$ be the corresponding measure with respect to $R_T$, then we can write the error of the options price estimated by our neural SDE $S_T$.
\begin{equation}
    \bigg| \mathbb{E}_W[f(S_T)] - \mathbb{E}_V[f(R_T)] \bigg|
\end{equation}
This is awfully close to Wasserstein distance of inequality (\ref{eq:nsde2}). Instead of just being a happy coincidence, we propose the following conjecture.
\begin{conj} \label{conj:1}
Given two Itô process $S_t$ and $R_t$ with with equivalent martingale measures $W$ and $V$ respectively and a function $f \in \mathcal{F}$ from the family of Lipshitz-1 functions, then this inequality holds. $$\bigg| \mathbb{E}_W[f(S_T)] - \mathbb{E}_V[f(R_T)] \bigg| \le \bigg| \mathbb{E}[f(S_T)] - \mathbb{E}[f(R_T)] \bigg|$$
\end{conj}
The proof of this conjecture in its most general form is by no means trivial and would constitute a new mathematical development as far as we know. A proof of a simplified version of this is given in the appendix based on the assumption that $S_t$ and $R_t$ are Brownian motion with drift, and not general processes. 

Lastly, let's us notice subtly what is going on here. By Girsanov (section \ref{sec_martingale}) we know that $$\mathbb{E}_W[f(S_T)]  = \mathbb{E} \left[f \left(S_t + \int_t^T \sigma_\phi(s,\omega)dB_s \right) \right]$$

and if $R_t = a(t,\omega)dt + b(t,\omega)dB_t$ then 
$$\mathbb{E}_V[f(R_T)]  = \mathbb{E} \left[f \left(R_t + \int_t^T b(s,\omega)dB_s \right) \right]$$
. We can now re-write the inequality from the conjecture, adding the computed Wasserstein distance. This helps reveal that when comparing options prices, we are only concerned with the stochastic integral part of the underlying process. In this light, the conjecture simply proposes that if the distributions of the underlying processes are close, then the distributions of their stochastic integral parts must be even closer.

$$\left| \mathbb{E} \left[f \left(S_t + \int_t^T \sigma_\phi(s,\omega)dB_s \right) \right] - \mathbb{E} \left[f \left(R_t + \int_t^T b(s,\omega)dB_s \right) \right] \right| \le W_1(S_T,R_T)$$

\section{Experiments}

The experiment was designed as a proof-of-concept in order to investigate the claims of the proposed method. In order to do this, empirical samples are drawn from a known process which has well-understood dynamics and a closed-form solution to the options pricing problem. This allows us to execute the pipeline and compare our results to the known solution, which wouldn't be available in practice. The process we consider is a geometric Brownian motion, for which the price of a European call option has a closed form solution given in \cite{bs}. 
    \begin{figure}
    \makebox[\textwidth][c]{
        \begin{tabular}{ccc}
          \includegraphics[height=25mm]{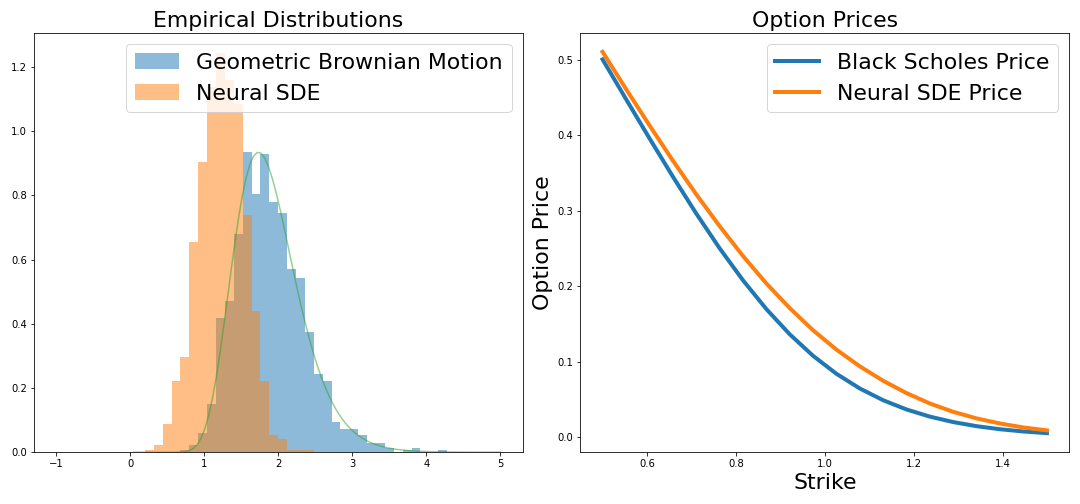} &   \includegraphics[height=25mm]{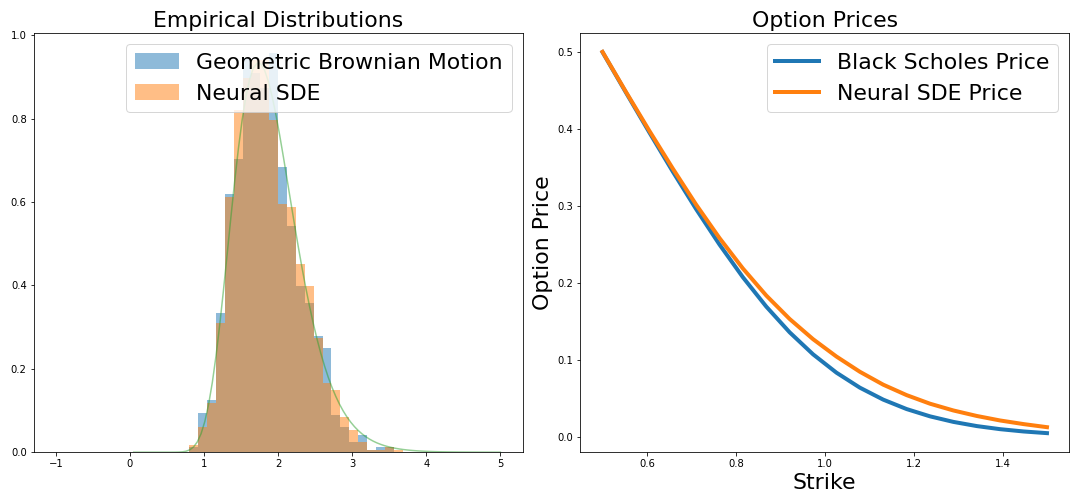} & 
          \includegraphics[height=25mm]{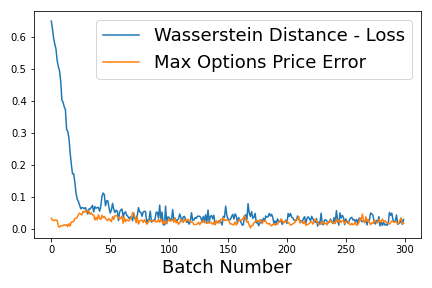}\\
         \includegraphics[height=25mm]{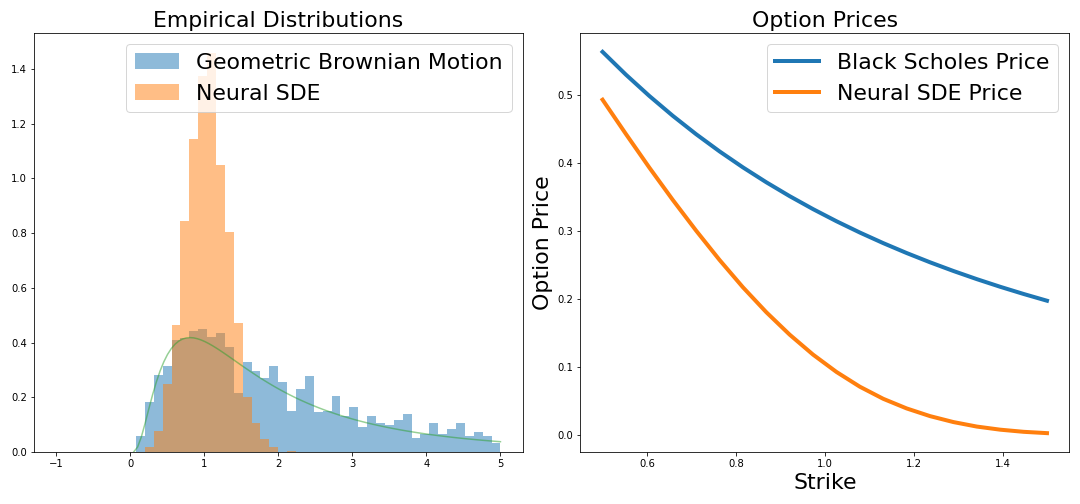} &   \includegraphics[height=25mm]{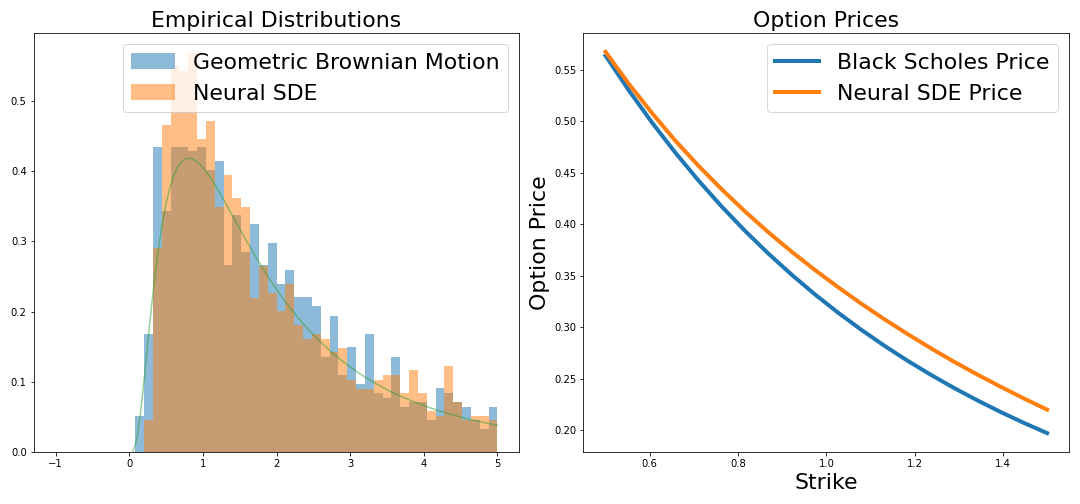} & 
          \includegraphics[height=25mm]{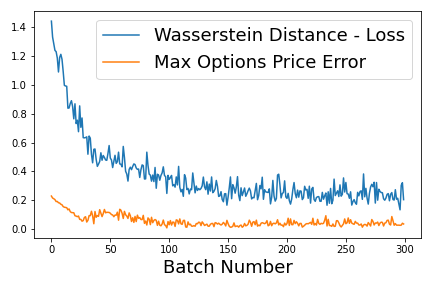} \\
         \includegraphics[height=25mm]{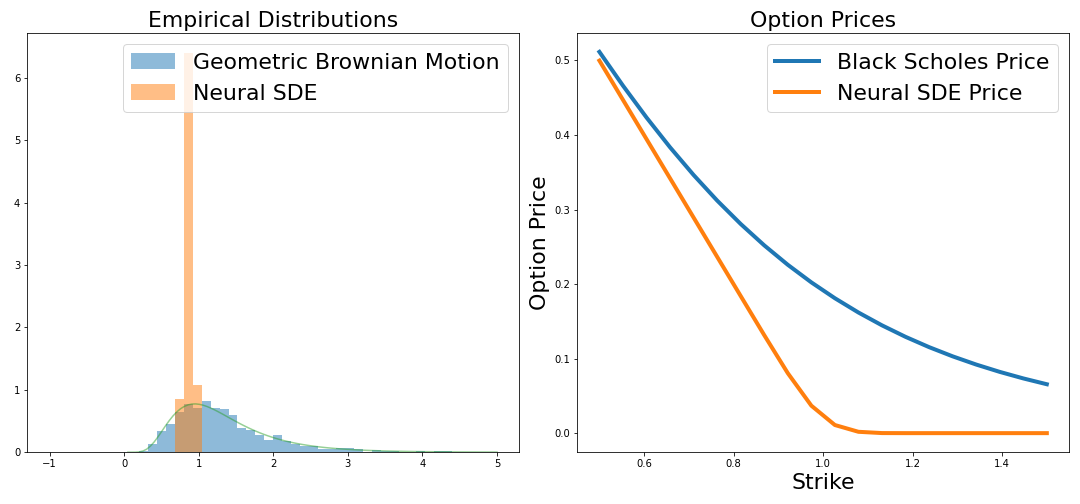} &   \includegraphics[height=25mm]{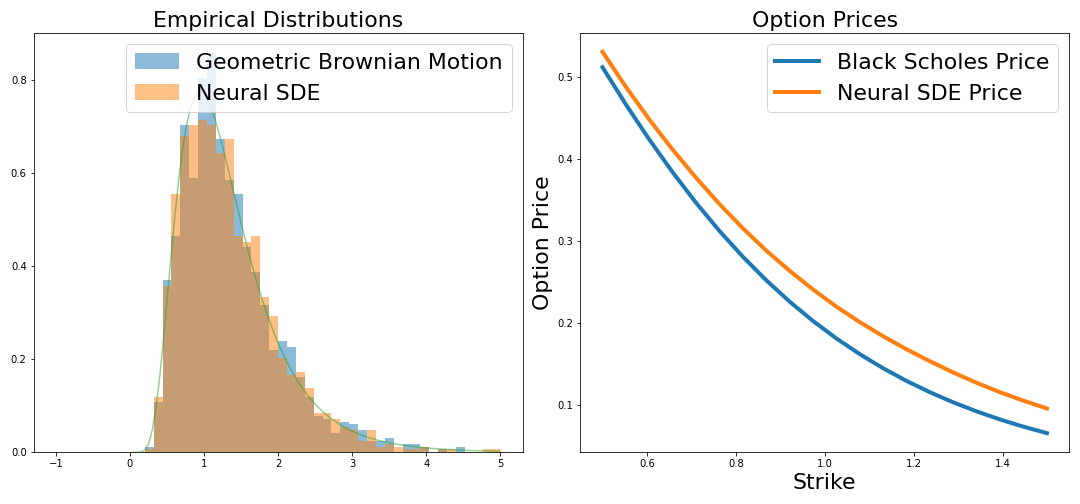} & 
          \includegraphics[height=25mm]{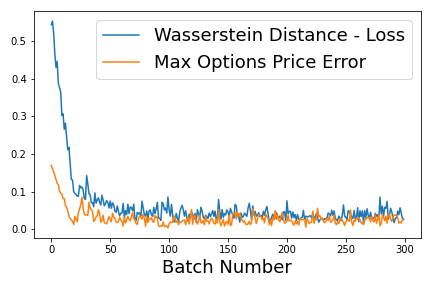} \\
        (a) After Initialization & (b) After Training  & (c) Loss - Error\\[6pt]
        \end{tabular}
    }
    \caption{In this figure are displayed the results of our main experiment. Each row corresponds to individual trials, where the only difference between them are the random values for the drift and diffusion constants of the target geometric Brownian motion process. (a) Shows the empirical distributions and options prices for several strikes of the neural SDE compared to the target before training has begun. This is contrasted with (b) which shows the same plots but after training. The batch that attained the lowest loss during training is displayed, which is a common practice. (c) Shows the loss compared to the maximum error in the options price over each batch during training. These two quantities are precisely those which are the subject of conjecture \ref{conj:1}.}
    \label{fig:1}
    \end{figure}
Under this experiment, we sample empirical data from $R_t$, which is defined by the SDE for geometric Brownian motion where $a$ and $b$ are constants that augment the drift and diffusion terms respectively.
\begin{equation}
    R_t = R_t( a\ dt + b\ dB_t )
\end{equation}

We then fit the neural SDE model, $S_t$, as defined in equation (\ref{eq:nsde}), where the drift and diffusion terms are 3-layer feed-forward neural networks with 100 nodes in each layer, Tanh activation functions and linear output. Sample trajectories are produced from $S_t$ by a numerical SDE solver. We used the \textit{sdeint} function from the torchsde package \cite{pmlr-v108-li20i,kidger2021neural}. The loss function from equation (\ref{eq:wass2}) is applied to the empirical distributions on batches of data from $S_t$ and $R_t$ at discrete time intervals throughout the trajectory. The total loss for the batch is then the average over these time steps. The loss function was implemented based on the algorithm described in \cite{ramdas2015wasserstein}. While we are effectively minimizing the same quantity as in the SDE-GAN architecture, we do so without playing the adversarial game with the discriminator. This approach works similarly to more traditional training or feed-forward or recurrent neural networks. It has shown to be a simple and effective way to train 1-dimensional neural SDEs.

Three random combinations of the drift and diffusion constants $a$ and $b$ are chosen and thus three trials are carried out. An additional experiment is performed as a proxy to investigate the claim of conjecture \ref{conj:1}. Here, we randomly initialize 1,000 pair of Neural SDEs and for each pair we compare the Wassterstein-1 distance between their distributions and the error in the options prices between them. These are the quantities corresponding to each side of the inequality in conjecture \ref{conj:1}.

The entire experiment is done in PyTorch. The only hyper-parameter that was tuned was the learning rate of the Adam optimizer. The entire experiment was performed on a linux server with 64 MB of memory and a Geforce RTX GPU. The compute time was about 6 hours. Example code will be made available online.

Results from the experiments positively support the claims of this research. Figure \ref{fig:1} first displays the empirical distributions of the Neural SDE models compared to the target measures before training on the left, and then after training on the right. For each trial the closed-form solution to the problem computed via the Black-Scholes formula is compared to the neural call option price for several strikes.  We find in each case our technique converges to an accurate fit. 

Moreover we investigate the bound proposed in conjecture \ref{conj:1} by plotting the loss value compared to the error of the options price during training of our models. As well we display the results of the stand-alone experiment in which we compare 1,000 random neural SDEs in figure \ref{fig:2}.

\section{Discussion}

Numerous studies have endeavored to solve the option pricing problem using deep neural networks and stochastic differential equations. \cite{chen2019deep} proposes a method to use deep neural networks to learn the dynamics of American options and optimizes based on minimizing the error over a backward SDE. A highlight here is efficient computation of options prices in high dimensions. Other models aim to utilized neural networks to learn option prices based on supervised learning \citep{Liu_2019,Brostrm2018ExoticDA} and claim these models accurately predict options prices. These methods contrast our research in that they require option price data for training while our approach does not, and hence can be considered an unsupervised method. 

This research is the first to directly implement neural SDEs for pricing options. Besides the main assumption that we can approximate any Itô process with a neural SDE, the rest of the computation follows directly from traditional mathematical finance theory. In this framework we do not need to make any assumptions on the form of the underlying price process like in traditional techniques nor do we need to supply expensive option data for supervised training. Under these considerations, the framework is a simple and natural extension of neural SDEs to mathematical finance. 

\begin{wrapfigure}{r}{0.33\textwidth}
    \makebox[0.33\textwidth][c]{
        \includegraphics[width=0.4\textwidth]{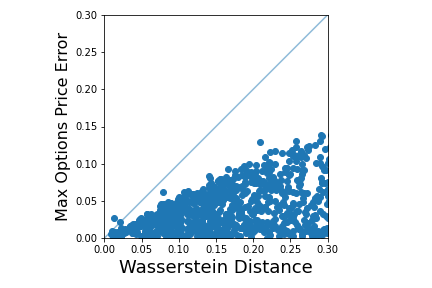}
    }
    \caption{In this experiment 1,000 pairs of neural SDEs where randomly initialized and compared in order to investigate the empirical trend between the Wasserstein distance of two distributions of processes and the errors of the implied options prices between them. This is a simple scatter plot. For reference we have added the line $x=y$. Any points plotted above this line would suggest the conjectured bound is not correct. }
    \label{fig:2}
\end{wrapfigure}

Given that this work is largely a proof-of-concept there remain limitations to implementing such a technique in a professional real-world setting. The first obstacle is that our current model can only handle 1-dimensional price data. This is largely due to the state-of-the-art in computing non-parametric estimates of the Wasserstein-1 distance in higher dimensions. In this paper we utilized the closed-form computation of \cite{ramdas2015wasserstein} which effectively compares CDFs, but there is an enticing method for computing the distance in higher dimensions via the dual optimization technique which will be studied further. Another way our technique could be extended to higher dimensions is by adapting the more general SDE-GAN approach, which theoretically works in higher dimensions and effectively computes the Wasserstein distance just as in our approach.

Finally, we acknowledge that the conjectured bound, if true, could be quite profound, however it seems unlikely it could be true in the most general sense. Indeed we see a couple data points stray across the boundary in Figure \ref{fig:2}. While it seems like an natural quantity to investigate, so far we haven't found existing results which shed light one way or the other on the matter. The motivation is that, if it were true, then we could guarantee a level of accuracy on the option price using minimal assumptions on the underlying price process. The accuracy of the options price would be bounded by our ability to fit the underlying data. This is particularly interesting when pricing options of assets for which we have little understanding of the underlying price process.

\section{Acknowledgements}

A big thanks to Ioannis Mitliagkas and Alexander Fribergh at Université de Montréal, and to Patrick Kidger at the University of Oxford for useful insights and support.

\appendix

\section{Appendix}
\subsection{Proof Sketch for the Proposed Bound - Brownian Motion with Drift}

The goal is to prove that for any two stochastic processes which can be represented as SDEs, $S_t$ and $R_t$, have options prices which are bounded by the Wasserstein-1 distance between them. This is to say that we are trying to prove the following inequality for all $T>0$.
\begin{equation}
    \bigg| E_W[f(S_T)] - E_V[f(R_T)] \bigg| \le \bigg| E[f(S_T)] - E[f(R_T)] \bigg|
\end{equation}

Where the notation $E_W[\cdot]$ means the expectation is taken over the equivalent martingale measure $W$ which turns $S_t$ into a martingale. Similarly $V$ is the measure which converts $R_t$ to a martingale. 

For this proof, we are going to simplify the assumptions, and only assume that $S_t$ and $R_t$ are special processes called Brownian motion with drift. Under this assumption, the SDEs are written as follows.

\begin{align}
    dS_t &= \mu dt + dB_t\\
    dR_t &= a dt + dB_t
\end{align}

Under this assumption, $\mu$ and $a$ are constants, and the diffusion function is the identity. These equations are called Brownian motion (BM) with drift, and the dynamics of these models are much better understood than general SDEs. In particular, the law of these processes is well understood. Let's take the case of a general BM w/ drift, $S_t$, like above. At any time $t \ge 0$, $S_t \sim \mathcal{N}(\mu t, t)$, a Gaussian normal random variable with mean $\mu t$ and variance $t$. This is basically all we need to prove our conjecture in this case.

Consider two processes, $S_t$ and $R_t$, as defined above and let $S_t \ne R_t$. This implies that $\mu \ne a$. Now let us consider the Wasserstei-1 distance between the empirical densities of $S_t$ and $R_t$ at some time $T>0$. We know that $S_T \sim \mathcal{N}(\mu T, T)$ and $R_T \sim \mathcal{N}(a T, T)$, and by the definition of distance metrics, we must have that $$W_1(S_T,R_T)>0$$
which implies that for all Lipschitz-1 or less functions, $f$, we have this.
$$\bigg| E[f(S_T)] - E[f(R_T)] \bigg|>0$$

Now let us consider the quantity that we are trying to bound.

$$\bigg| E_W[f(S_T)] - E_V[f(R_T)] \bigg| = \bigg| E_W \left[f \left( \int_0^T \mu dt + \int_0^T dB_t \right) \right] - E_V\left[f \left(\int_0^T a dt + \int_0^T dB_t\right) \right] \bigg|$$

$W$ and $V$ are the equivalent martingale measures implied by Girsanov theory \citep{girsanov}. According to the theory, under the measures $W$ and $V$, $B_t$ is no longer a BM, but instead, BM is different under each measure. We make use of the following lemma which is a standard result of Girsanov theory.

\begin{lemma}
For a general drift-diffusion process $X_t$ defined as $$dX_t = \mu(t,\omega) dt + \sigma(t, \omega) dB_t$$ which is adapted to the Brownian filtration. We denote the nature probability measure $P$, then $$d\tilde{B}_t = dB_t + \frac{\mu(t,\omega)}{\sigma(t, \omega)}dt$$
defines a Brownian motion under the equivalent martingale measure $Q$. Under $Q$
$$dX_t = \sigma(t,\omega)d\tilde{B}t$$
\end{lemma}

Using this lemma, under $W$, BM will be written as $B^W_t$, like this
\begin{align}
    B^W_t = \int_0^t dB_t + \int_0^t \mu ds
\end{align}
and similarly, under $V$, BM will be writte as $B^V_t$, as follows.
\begin{align}
    B^V_t = \int_0^t dB_t + \int_0^t a ds
\end{align}

We should all have noticed that under the measure $W$, $S_t$ is a BM and under $V$, $R_t$ is a BM. Therefore, in order to compute those expectations over $W$ and $V$, we substitute in the BMs under each measure and compute the expectations back in the natural measure.
\begin{align}
    E_W \left[f \left( \int_0^T \mu dt + \int_0^T dB_t \right) \right] &= E_W \left[f \left( B^W_T \right) \right] \\
    &= E \left[f \left( B_T \right) \right]
\end{align}
and 
\begin{align}
    E_V \left[f \left( \int_0^T a dt + \int_0^T dB_t \right) \right] &= E_V \left[f \left( B^V_T \right) \right] \\
    &= E \left[f \left( B_T \right) \right]
\end{align}

And now we summarize our steps until now.

\begin{align}
\bigg| E_W[f(S_T)] - E_V[f(R_T)] \bigg| &= \bigg| E[f(B_T)] - E[f(B_T)] \bigg| \\
&= 0 
\end{align}

We certainly now have proven the desired bound for this case:

\begin{align}
    \bigg| E_W[f(S_T)] - E_V[f(R_T)] \bigg| \le \bigg| E[f(S_T)] - E[f(R_T)] \bigg|
\end{align}

This case is somewhat trivial in that BM with drift is always converted to the same law under the transformation to the equivalent martingale measure, so the distance between any two Brownian motions with drift will be zero under the martingale measures. Proof of the claim for general drift-diffusion Itô processes would be much more profound and is the topic of current study.

\bibliographystyle{abbrvnat}
\bibliography{mainbib}

\end{document}